\documentclass{pasa}%

\newcommand{\p}{\partial}

\newcommand{\ie}{i.e. }
\newcommand{\eg}{e.g. }
\newcommand{\etal}{et al. }

\title[The explosion mechanism of core-collapse supernovae]{The explosion mechanism of core-collapse supernovae:\\
progress in supernova theory and experiments}
\author[T. Foglizzo]{Thierry Foglizzo$^1$, R\'emi Kazeroni$^1$, J\'er\^ome Guilet$^2$, Fr\'ed\'eric Masset$^3$, Matthias Gonz\'alez$^1$, 
Brendan K. Krueger$^{1,}$\thanks{present address: Los Alamos National Laboratory, NM 87545, USA}, 
J\'er\^ome Novak$^4$, Micaela Oertel$^4$, J\'er\^ome Margueron$^5$, Julien Faure$^1$, No\"el Martin$^6$, Patrick Blottiau$^7$, Bruno Peres$^8$ \and Gilles Durand$^1$\\
\affil{$^1$Laboratoire AIM (CEA/Irfu, CNRS/INSU, Univ. Paris Diderot), CEA Saclay, F-91191 Gif sur Yvette, Cedex, France}%
\affil{$^2$Max Planck Institute for Astrophysics, Karl-Schwarzschild-Str. 1, 85748 Garching, Germany}%
\affil{$^3$Instituto de Ciencias F\'isicas, Universidad Nacional Autonoma de M\'exico, P.O. Box 48-3, 62251 Cuernavaca, Morelos, Mexico}%
\affil{$^4$LUTh, CNRS/Observatoire de Paris/Univ. Paris Diderot, 5 place Jules Janssen, F-92195 Meudon, France}%
\affil{$^5$Institut de Physique Nucl\'eaire de Lyon, Univ. Claude Bernard Lyon 1, IN2P3-CNRS, F-69622 Villeurbanne, France}%
\affil{$^6$Institut de Physique Nucl\'eaire, IN2P3-CNRS, Univ. Paris-Sud, F-91406 Orsay cedex, France}%
\affil{$^7$CEA, DAM, DIF, F-91297 Arpajon, France}%
\affil{$^8$Dept d'Astronomia i Astrof\'isica, Univ. de Valencia, Edifici d'Investigaci\'o J. Munyoz, C/ Dr. Moliner, 50, 46100 Burjassot, Spain}}%
\jid{PASA}
\doi{10.1017/pas.\the\year.xxx}
\jyear{\the\year}

\begin{document}%
\begin{abstract}
The explosion of core-collapse supernova depends on a sequence of events taking place in less than a second in a region of a few hundred kilometers at the center of a supergiant star, after the stellar core approaches the Chandrasekhar mass and collapses into a proto-neutron star, and before a shock wave is launched across the stellar envelope. Theoretical efforts to understand stellar death focus on the mechanism which transforms the collapse into an explosion. Progress in understanding this mechanism is reviewed with particular attention to its asymmetric character. We highlight a series of successful studies connecting observations of supernova remnants and pulsars properties to the theory of core-collapse using  numerical simulations. The encouraging results from first principles models in axisymmetric simulations is tempered by new puzzles in 3D.
The diversity of explosion paths and the dependence on the pre-collapse stellar structure is stressed, as well as the need to gain a better understanding of hydrodynamical and MHD instabilities such as SASI and neutrino-driven convection. The shallow water analogy of shock dynamics is presented as a comparative system where buoyancy effects are absent. 
This dynamical system can be studied numerically and also experimentally with a water fountain.
The potential of this complementary research tool for supernova theory is analyzed. We also review its potential for
public outreach in science museums.
\end{abstract}
\begin{keywords}
accretion -- hydrodynamics -- instabilities -- shock waves -- supernovae
\end{keywords}
\maketitle%
\section{INTRODUCTION}

The explosive death of massive stars is a key ingredient in stellar evolution, stellar population synthesis and the chemical enrichment of galaxies. It defines the birth conditions of neutron stars which, if associated in a coalescing binary system, may be responsible for short GRBs and r-process nucleosynthesis. The shock wave launched in the interstellar medium during a supernova explosion accelerates cosmic rays and may contribute to triggering star formation. Understanding the mechanism of supernovae explosions has become a priority since the detailed observation of SN1987A neutrinos and the identification of its massive progenitor. It is still a theoretical challenge despite hundreds of new events observed every year in distant galaxies, and the progress of computational power. Even if robust explosions are not obtained from first principles yet, numerical simulations are able to explore physical ideas in full 3D models, filling the gap between theoretical concepts and observations. The multiplication of theoretical results may seem difficult to interpret in view of the multiplicity of physical assumptions, which range from simple adiabatic approximations of an ideal gas to advanced modeling of neutrino transport and nuclear interactions in general relativity. Substantial progress has been achieved  over the last ten years with an emphasis on the asymmetric character of the explosion, which we propose to summarize in the present review. \\
Some important observational constraints regarding the asymmetric character of the explosion are recalled in Sect.~\ref{sect_asym}. The general theoretical framework of neutrino-driven explosions is summarized in Sect.~\ref{sect_framework}. In order to underline the nature of recent theoretical progress we have shown in Sect.~\ref{sect_before2009} that a large set of modern physical ideas were already known before 2009 and later confirmed by more advanced calculations. We analyze in Sect.~\ref{sect_after2009} the new ideas which have driven supernova theory in the most recent years beyond the production of better tuned models. In particular the diversity of explosion scenarios has drawn attention to the pre-collapse stellar structure and correspondingly increased the size of the parameter space of initial conditions to be explored. This complexity calls for a deeper understanding of the physical processes. An experimental fountain based on a shallow water analogy has been proposed to gain insight into one of the instabilities responsible for the asymmetric character of the explosion. The potential of this new tool is analyzed in Sect.~\ref{sect_fountain}, both for theoretical research and for public outreach. The study of neutrino and gravitational wave diagnostics are left to the recent reviews of Janka \etal (2012), Janka (2012), Kotake \etal (2012), Burrows (2013) and Kotake (2013). Particular attention is paid to the many new results discovered over the last two years since these reviews.

\section{OBSERVATIONAL EVIDENCE FOR ASYMMETRIC EXPLOSIONS \label{sect_asym}}

The observed distribution of pulsar velocities (\eg Hobbs \etal 2005) has been a puzzle for more than 40 years because they are much faster than massive stars (Gunn \& Ostriker 1970).  Average velocities of several hundreds of km/s cannot be explained by a residual orbital velocity gained through the disruption of a binary system during the explosion. A pulsar kick is a natural outcome of an asymmetric explosion process with a significant $l=1$ component as seen in Sect.~\ref{sect_kick}. The angle between the direction of the kick and the direction of the rotation axis of the pulsar contains a very interesting constraint on the geometry of the explosion. The direction of the rotation axis can be accurately determined when a pulsar wind nebula is observed, but this sample is still small and lacks unambiguous cases where the kick is strong enough to neglect  a possible orbital contribution to the kick (Ng \& Romani 2004, Wang \etal 2006). Polarimetry suggests a strong correlation but cannot fully disentangle kick-spin alignment from orthogonality (Noutsos \etal 2012). This correlation is smeared out by the Galactic potential for pulsars older than 10 Myr (Noutsos et al. 2013).  \\
The asymmetric character of the explosion is also suggested by the sudden increase of polarization of the light observed from a type II-P supernova explosion after $\sim$90 days, at the moment of transition from the photospheric phase to the nebular phase (Leonard \etal 2006). This increase coincides with the moment when the innermost regions become visible.\\
On a longer timescale, the shape of the inner ejecta currently seen in SN1987A also suggests an asymmetric explosion geometry which is not aligned with the large scale structure of the circumstellar medium (Larsson \etal 2013). Even three hundred years after the explosion, its asymmetric character can leave an imprint on the chemical composition of the ejecta. The spatial distribution of $^{44}$Ti observed by the NuSTAR space telescope in Cassiopea A shows a global asymmetry, which seems to favor the region opposed to the direction of the compact object (Grefenstette \etal 2014). This one-sided asymmetry seems to confirm the theoretical prediction of Wongwathanarat \etal (2013) where a strong one-sided shock expansion is favorable to nucleosynthesis while the closer region in the opposite direction attracts the neutron star gravitationally. \\
Direct constraints on the first second of the explosion are expected from the future detection of neutrinos (\eg Wurm et al. 2012, M\"uller \etal 2014a) and gravitational waves (Ott 2009, Kotake 2013, M\"uller \etal 2013) from a Galactic supernova. The future detection of the diffuse neutrino background will globally constrain both the physics of the explosion and the supernova rate (Beacom 2010). For an individual Galactic event, the time variability of the neutrino signal from the IceCube experiment could directly measure the modulation of the emission due to a SASI oscillation of the shock (Lund et al. 2010, 2012, Tamborra et al. 2013, 2014).

\section {THEORETICAL FRAMEWORK\label{sect_framework}}

\subsection{Neutrino-driven explosion scenario}

\begin{figure}
\begin{center}
\includegraphics[width=0.7\columnwidth]{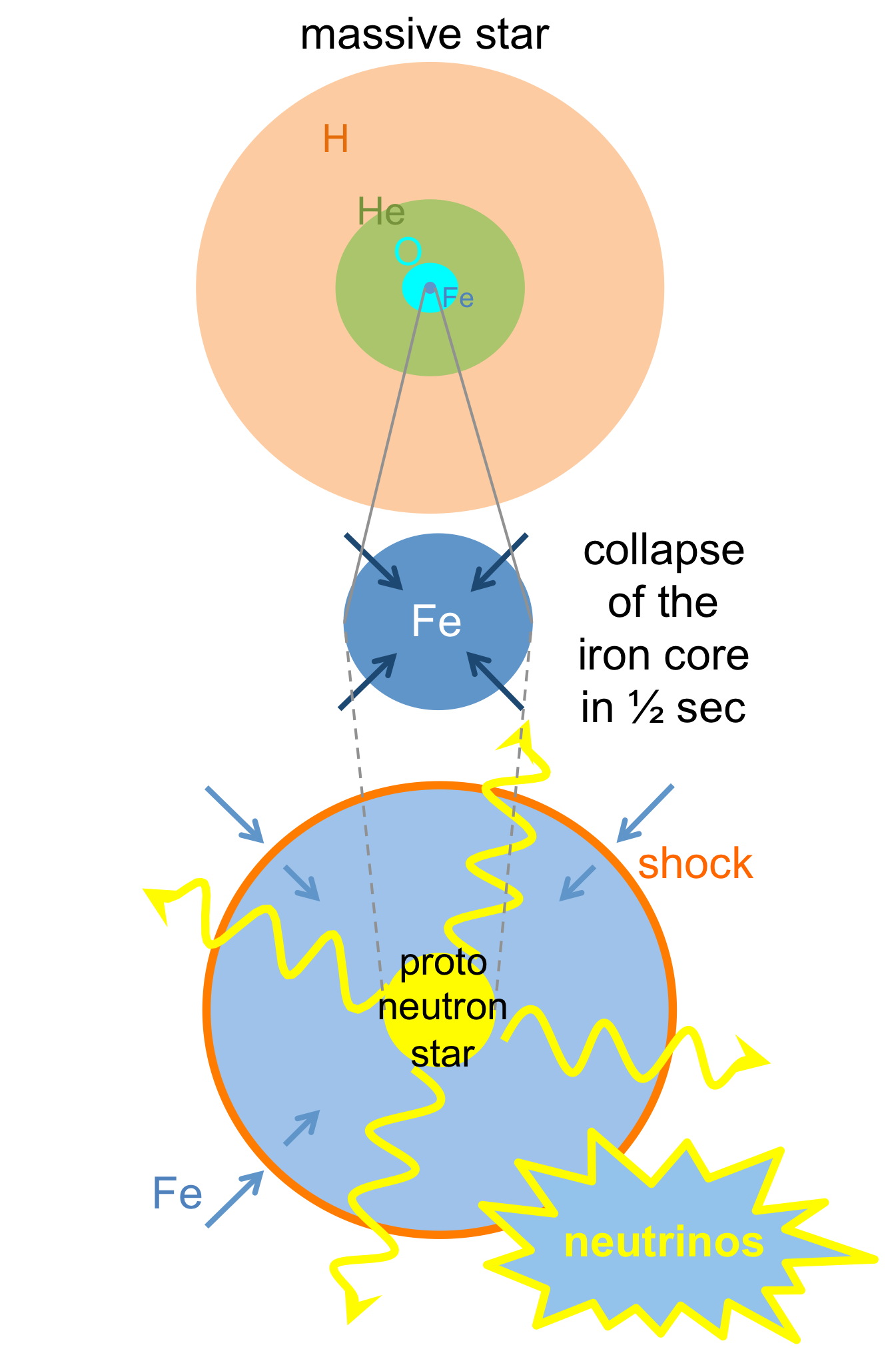}
\caption{The neutrino-driven delayed explosion mechanism relies on the absorption of neutrinos by the dense post-shock gas.}\label{Fig_collapse}
\end{center}
\end{figure}

The most promising framework to understand the majority of supernova explosions was set by Bethe \& Wilson (1985). The delayed explosion mechanism driven by neutrino energy was initially described as a spherically symmetric process. As the mass of the central core approaches the Chandrasekhar mass ($\sim 1.4M_{\rm sol}$), the pressure of degenerate relativistic electrons becomes insufficient to resist gravity. Electron pressure is further decreased as they are captured by protons and produce neutrons and neutrinos. A new equilibrium is reached between gravity and nuclear matter essentially made of neutrons. The stellar core collapses from a radius of $\sim1500$km to this new equilibrium of a few tens of kilometers in less than half a second (Fig.~\ref{Fig_collapse}). 
The gravitational energy gained from this collapse seems large enough to be able to account for the observed kinetic energy $\sim 10^{51}{\rm erg}$ of the supernova ejecta:
\begin{eqnarray}
{GM_{\rm ns}^2\over R_{\rm ns}}\sim 2\times 10^{53}{\rm erg}\left({30{\rm km}\over R_{\rm ns}}\right)
\left({M_{\rm ns}\over 1.5M_{\rm sol}}\right)^2.
\end{eqnarray}
A major difficulty of supernova theory is to explain how this gravitational energy is transferred to the stellar enveloppe to reverse the infalling motion into an explosion. The infall of supersonic matter onto the surface of this proto-neutron star produces a deceleration shock which stalls at a radius of $\sim150$km. The bounce of the free-falling matter $v_{\rm ff}^2/2\sim GM/r$ onto the core is far from elastic, because a significant fraction of the kinetic energy is absorbed into the dissociation of iron nuclei. This can be seen by comparing the kinetic energy of a free-falling nucleon and its binding energy 8.8MeV in the iron atom, neglecting special relativistic corrections for the sake of simplicity:
\begin{eqnarray}
{{1\over 2}m_{\rm n}v_{\rm ff}^2\over 8.8{\rm MeV}}
&\sim& {236 {\rm km}\over r}
{M\over 1.5 M_{\rm sol}}.
\end{eqnarray}
Stellar matter passing through the shock is thus dissociated into free nucleons. The subsonic advection of these nucleons towards the surface of the proto-neutron star takes place in an intense flux of neutrinos which diffuse out of the proto-neutron star and carry away most of the gravitational energy gained during the contraction. The detection of neutrinos from SN1987A were instrumental to confirm the premises of supernova theory set by Colgate \& White (1966).\\
The post-shock region is made of two successive regions defined by the direction of the reaction 
\begin{eqnarray}
p+e\leftrightarrow n+\nu,
\end{eqnarray}
where the relativistic velocity of the electrons corresponds to a Lorentz factor $\gamma_e >({m_{\rm n}-m_{\rm p})/ m_{\rm e}}\sim 2.5$.\\
Although most neutrinos emerge from the neutrinosphere of the cooling neutron star, a fraction comes from the neutronization of accreted matter below the gain radius. This process decreases the entropy of neutron rich matter settling in a stably stratified manner. \\
Between the shock surface and the gain radius, the dominance of neutrino absorption over neutrino emission increases the entropy of the matter. This gain region is crucial to the success of the explosion, which relies on the absorption of sufficient neutrino energy to revive the stalled shock.\\
Other sources of energy have been considered to produce an explosion, such as the rotational and the magnetic energies. The rotation period of pulsars at birth defines a reference rotational energy $E_{\rm rot}$ deduced from the conservation of angular momentum
\begin{eqnarray}
E_{\rm rot}&\sim& 2.8\times10^{50}{\rm erg}\left({M_{\rm ns}\over 1.4 M_{\rm sol}}\right)
\left({R_{\rm ns}\over10{\rm km}}\right)^2\left({10{\rm ms}\over P_{\rm ns}}\right)^2.\label{Erot}
\end{eqnarray}
Spin periods $P_{\rm ns}\sim10-20$ms at birth are considered by Heger \etal (2005) as plausible extrapolations from the observations of young pulsars, without excluding possible effects of binary interaction during the lifetime of the massive star (Sana \etal 2012, De Mink \etal 2013, 2014) or a significant redistribution of angular momentum during or immediately after the pulsar birth. The relative inefficiency of known spin-down mechanisms (Ott \etal 2006) rules out rotationally-driven supernovae as the generic case. According to Eq.~(\ref{Erot}), the rotational energy could be a major contributor in the particular cases where the rotation rate of the stellar core is large enough to produce millisecond pulsars.  \\
Magnetic energy could also play an important role if the magnetic field within the core were strong enough. The main source of magnetic field amplification is the differential rotation within the core (\eg Akiyama \etal 2003), which is only a fraction of the total rotational energy. \\
Assuming that the rotational energy is too weak to be the dominant contributor of the most common explosions (\ie $P_{\rm ns}\ge 10$ms), most studies of core collapse have focused on the challenge of producing an explosion without any rotation at all.
We shall see in Sect.~\ref{sect_spin} that rotation could be an important ingredient of the explosion mechanism even if its total energy is modest. \\
We can also expect magnetic effects to play a role in shaping the geometry of the explosion even if the magnetic energy is modest compared to the final kinetic energy of the ejecta (\eg Guilet \etal 2011).\\

Detailed numerical modeling solving the Boltzmann equation to describe neutrino transport and taking into account special and general relativistic effects reached the conclusion that the scenario proposed by Bethe \& Wilson (1985) is unable to power the spherically symmetric explosion of massive stars (Liebendoerfer \etal 2001) except for the lightest ones. The envelope of stars in the range 8-10 $M_{\rm sol}$ is light enough to allow for a spherical explosion of $\sim 10^{50}$erg in about 100ms (Kitaura \etal 2006).
In a sense, the absence of systematic explosions in spherical symmetry is consistent with the observational evidence summarized in Sect.~\ref{sect_asym}.
Theoretical efforts over the past two decades have updated the delayed neutrino-driven explosion scenario by taking into account its multidimensional nature. 

\subsection{Hydrodynamical sources of asymmetry}

\subsubsection{Neutrino-driven buoyancy}

In addition to the prompt convection associated with the deceleration of the shock as it stalls, the continued heating of the gas by neutrino absorption maintains a radial entropy gradient oriented in the same direction as gravity from the shock to the gain radius (Herant \etal 1992, Janka \& M\"uller 1996). Some gravitational energy can be gained if the flow is able to interchange high and low entropy layers. The Brunt-V\"ais\"al\"a frequency $\omega_{\rm BV}$ characterizes the timescale of the fastest motions fed by buoyancy forces. Defining the entropy $S$ in a dimensionless manner, the Brunt-V\"ais\"al\"a frequency is expressed as follows:
\begin{eqnarray}
S&\equiv&{1\over\gamma-1}\log {P/P_0\over(\rho/\rho_0)^\gamma},\\
\omega_{\rm BV}^2&\sim& -{\gamma-1\over\gamma}\nabla S\cdot\nabla\Phi.
 \end{eqnarray}
This interchange can feed turbulent motions and push the shock further out, increasing the size of the gain region and diminishing the energy losses due to dissociation. Numerical simulations performed in the early 2000's were disappointing though (Buras \etal 2003), both because the duration of the simulation was limited to a few hundred milliseconds and as some equatorial symmetry was assumed due to limited computational resources. The equatorial symmetry precluded the growth of global $l=1$ modes.
Foglizzo et al. (2006) pointed out that the negative sign of the entropy gradient is not a sufficient criterion for the convective instability in the gain region because the interchange has to be fast enough to take place before the advected gas reaches the gain radius. A criterion for linear stability compares the advection timescale to the buoyancy timescale estimated from the Brunt-V\"ais\"al\"a growth rate:
\begin{eqnarray}
\chi\equiv \int_{\rm gain}^{\rm shock} |\omega_{\rm BV}|{{\rm d} r\over |v_r|}<3
\end{eqnarray}
From this criterion one can anticipate that the strength and consequences of neutrino-driven buoyancy may vary from one progenitor to another, depending on the radius of the stalled shock.

\subsubsection{Instability of the stationary shock: SASI}

\begin{figure}
\begin{center}
\includegraphics[width=0.7\columnwidth]{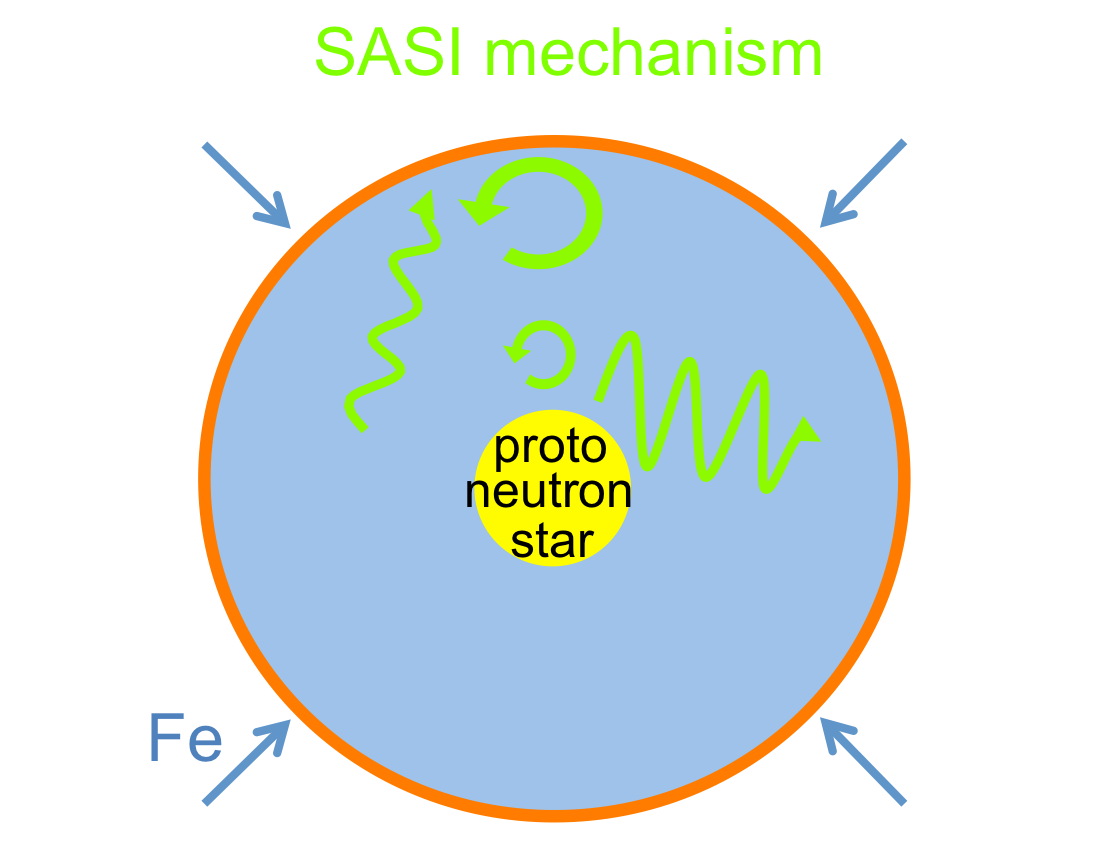}
\caption{SASI mechanism based on the coupling between acoustic waves (wavy arrows) and advected perturbations (circular arrows) between the stalled shock and the proto neutron star.}\label{Fig_scheme}
\end{center}
\end{figure}

Another source of symmetry breaking was discovered by Blondin \etal (2003) who studied a simplified setup where neutrino heating was neglected in order to avoid any possible confusion with neutrino-driven convection. The Standing Accretion Shock Instability (SASI) corresponds to a global oscillatory motion of the shock surface with a period comparable to the advection time from the shock to the neutron star surface. This global ($l=1$) and oscillatory character are distinct features of the linear regime when perturbations grow exponentially with time. In contrast, the convective instability is non-oscillatory and dominated by smaller azimuthal scales (typically $l=5-6$) comparable to the radial size of the gain region.\\
The mechanism responsible for SASI relies on the interaction of acoustic perturbations and advected ones, such as entropy and vorticity perturbations (Galletti \& Foglizzo 2005, Ohnishi 2006, Foglizzo \etal 2007, Scheck \etal 2008, Guilet \& Foglizzo 2012). As illustrated in Fig.~\ref{Fig_scheme}, a perturbation of the shock surface produces entropy and vorticity perturbations which are advected towards the proto-neutron star surface. Their deceleration in the hot region close to the neutron star produces an acoustic feedback which propagates towards the shock surface, pushes it and regenerates new advected perturbations with a larger amplitude than the first ones. This advective-acoustic cycle has been described analytically in a Cartesian geometry (Foglizzo 2009, Sato et al. 2009). The same coupling mechanism is responsible for the instability of the bow shock in Bondi-Hoyle-Lyttleton accretion (Foglizzo 2001, 2002, Foglizzo \etal 2005).

\section{WHAT DID WE KNOW BACK IN 2009?\label{sect_before2009}}

\subsection{First successful explosions in numerical simulations from first principles}

The discovery of the global $l=1$ oscillations of SASI emphasized the importance of performing numerical simulations over the full sphere. Using a ray by ray approximation of neutrino transport and relativistic corrections to the gravitational potential, Marek \& Janka (2009) obtained the first successful explosions of 11.2 and 15 $M_{\rm sol}$ progenitor models from first principles. The explosion was more difficult to obtain with the more massive progenitor due to the intense ram pressure of infalling matter. It took 800ms of post-bounce evolution (and three years of computation) for the axisymmetric shock to reach an explosion after a long phase of SASI oscillations. The easier explosion of 2D simulations compared to radial ones was analyzed by Murphy \& Burrows (2008) who showed that post-shock gas trapped in convective patterns is exposed for a longer time to the neutrino flux, thus accumulating enough energy and entropy in the gain region to push the shock outwards. The turbulent motions induced by hydrodynamical instabilities also contribute to the shock revival by building up a turbulent pressure in the post-shock region as observed by Burrows \etal (1995) and more recently stressed by Murphy \etal (2013) and Couch \& Ott (2014).

\subsection{Instabilities in the supernova core as an explanation for observed asymmetries?\label{sect_asym}}

\subsubsection{The mystery of pulsar kicks potentially explained\label{sect_kick}}

The Garching team was first to recognize the potential of global $l=1$ deformations to explain pulsar velocities of several hundreds of kilometers per second (Scheck \etal 2004). The global conservation of linear momentum implies that the final momentum of the neutron star is equal in magnitude and opposed in direction to the total momentum of the ejecta. 
The asymmetric distribution of ejected matter has been studied in a series of axisymmetric simulations where the neutrino luminosity was adjusted to trigger an explosion. They showed that the gravitational interaction pulls the neutron star in the direction of  the closest dense regions of post-shock gas, over a timescale of several seconds which can exceed the timescale of direct advection of momentum from the accreted gas. Their statistical study produced a shape of the distribution of pulsar velocities in agreement with observational constraints (Scheck \etal 2006). The magnitude of the kick is a stochastic variable, while the explosion energies and timescales were found to be deterministic in this study. The magnitude of the kicks obtained has been subsequently  confirmed with a different axisymmetric code by Nordhaus \etal (2010a, 2012) which directly followed the motion of the neutron star. This kick mechanism has been confirmed in 3D simulations by Wongwathanarat \etal (2010, 2013) who used an axis-free Yin-Yang grid rather than spherical coordinates.

\subsubsection{Enhanced mixing triggered by asymmetric shock in SN1987A}

The light curve of SN1987A, together with the early emergence of X-ray and gamma rays suggested that significant mixing disrupted the onion like structure of the star (McCray 1993, Utrobin 2004, Fassia \& Meikle 1999), probably triggered by the interaction of the shock with composition interfaces. The efficiency of this mixing is enhanced if the shape of the shock is non-spherical, as shown by Kifonidis \etal (2006). This study considered shock deformations dominated by the modes $l=1,2$ as naturally produced by SASI during the first second after shock bounce. The subsequent hydrodynamical evolution of the shock across the stellar envelope up to the surface showed final iron group velocities up to 3300km/s, strong mixing at the H/He interface, and hydrogen mixed down to velocities of 500km/s. These features are closer to observations than those obtained by the same team from a spherical shock dominated by the smaller scale convective instability (Kifonidis \etal 2003). The viability of this process in 3D was later confirmed by Hammer \etal (2010), who followed the evolution of the shock calculated in 3D by Scheck (2007), across the stellar enveloppe. They found that the development of mixing instabilities is more efficient in 3D and allows for higher clump velocities due to the different action of drag forces. The effect of the progenitor structure was recently explored by Wongwathanarat \etal (2014) with a series of red and blue supergiants of $15M_{\rm sol}$ and $20M_{\rm sol}$.

\subsection{First hints of a dominant spiral mode in 3D and the spin rate of neutron stars\label{sect_spin}}

As early as 2007, the first 3D simulations of the SASI instability were performed in the simplest adiabatic approximation (Blondin \& Mezzacappa 2007).  Instead of the sloshing oscillation $l=1$, $m=0$ ubiquitous in 2D axisymmetric simulations, the 3D simulations revealed a dominant spiral mode $l=1$, $m=\pm1$ even though no angular momentum was contained in the free falling stellar matter. This opened the possibility that a non-rotating stellar core could give birth to a rotating neutron star, with the opposite angular momentum carried away by the ejecta of the supernova. Blondin \& Mezzacappa (2007) estimated that the final spin period of the neutron star could be as short as 50ms. They checked that these dynamics were not an artifact of the inflowing inner boundary condition by performing equatorial simulations with a hard surface and a cooling function (Blondin \& Shaw 2007). \\
Another spectacular result was obtained by Blondin \& Mezzacappa (2007) by considering for the first time the 3D development of SASI in a rotating progenitor. They observed that the spiral mode is preferentially triggered in the same direction as the progenitor rotation, and as a consequence the angular momentum received by the proto-neutron star is opposed to that of the parent star. The phase of SASI instability may thus decelerate a neutron star to a lower spin than naively deduced from the conservation of angular momentum. They showed that a rotating massive star may even give birth to a counter rotating neutron star. These results were confirmed by Yamasaki \& Foglizzo (2008) using a perturbative approach in a cylindrical geometry. In particular the growth rate of SASI was shown to depend linearly on the angular momentum of the progenitor. Rotation is able to destabilize the prograde mode of SASI through the difference of rotation rates between the shock radius and the inner radius, even when the rotation rate is so small that the centrifugal force is negligible.\\
The existence of a physical mechanism capable of modifying the angular momentum of the neutron star during the collapse revived the debate about the comparison between the angular momentum in the stellar core inferred from stellar evolution and the angular momentum of pulsars at birth inferred from observations.
The study of Heger \etal (2005) had concluded that the spin rates of 10-15ms of the most common pulsars were compatible with stellar evolution with plausible magnetic field strengths deduced from dynamo action, without the need to invoke any magnetic braking during or after the explosion. Alternate scenarios are possible though, since the transport of angular momentum in the 1-D codes of stellar evolution relies on uncertain prescriptions based on debated dynamo action (Spruit 2002, Zahn \etal 2007). In particular the distribution of angular momentum could also be affected by the action of internal gravity waves (\eg Lee \& Saio 1993, Talon \& Charbonnel 2003, Pantillon \etal 2007, Lee \etal 2014). 

\subsection{The uncertainties of a non-axisymmetric explosion scenario}

The non-axisymmetric shock geometry observed in the first 3D simulations by Blondin \& Mezzacappa (2007) cast doubts on all the conclusions drawn in the axisymmetric hypothesis and urged new 3D simulations incorporating non-adiabatic processes. Would the third dimension be a key ingredient for robust successful explosions? Contrasting with the ample deformations of the shock found in the idealized simulations of Blondin \& Mezzacappa (2007), the first non-adiabatic simulations by Iwakami \etal (2008) suggested that the asymmetries induced by SASI may be weaker in 3D than in 2D. Back in 2009 it was very unclear whether the pulsar kick would reach a comparable magnitude in 3D and in 2D, and whether the efficiency of mixing induced by the shock propagation in the stellar enveloppe would compare as favorably to SN1987A in 3D simulations as in 2D. As already noted in Sect.~\ref{sect_asym}, these two questions later received positive answers. Launching an explosion from first principles, however, turned out to be even more challenging in 3D than in 2D.  

\section{WHAT IS NEW SINCE 2009?\label{sect_after2009}}

The progress of supernova theory in the last 5 years has gone much further than confirming previous results with improved computational power. New constraints have been obtained from the determination of neutron star masses. New ideas have emerged concerning the diversity of explosion paths and their sensitivity to the structure of the progenitor above the iron core including its angular momentum, magnetic field and initial asymmetries.

\subsection{A better constrained equation of state}

The equation of state of neutron rich matter at nuclear densities is an
important ingredient of the dynamics of core-collapse, which sets the radius
of the proto-neutron star and thus the depth of the gravitational potential.
Most simulations have used a parametrized equation of state provided by
Lattimer \& Swesty (1991), Hillebrandt \& Wolff (1985) or Shen et
al. (1998). The description of matter properties at nuclear densities is
extrapolated from the properties of nuclei available in terrestrial
experiments, as well as measurements of the mass and to a lesser extent the
radius of neutron stars.  The measurement of the mass of a neutron star with a
mass of $1.97 \pm 0.04 M_{\rm sol}$ by Demorest \etal (2010) is accurate enough
to rule out a series of alternate models and significantly reduces the
uncertainty associated to the equation of state. A second neutron star in this
mass range has been discovered by Antoniadis \etal (2013), with a mass of
$2.01 \pm 0.04 M_{\rm sol}$ for the pulsar PSR J0348+0432. Exotic forms involving
a transition to quark matter (Sagert \etal 2009, Fischer \etal 2011) can produce
significant dynamical consequences during core collapse, but involve phenomenological 
models with parameters adjusted to reproduce the observed massive neutron stars.
In particular, the properties of quark matter at the relevant densities are neither 
accessible to experiments nor to any theoretical
development from first principles.\\
Staying in the classical framework, some improvements are
necessary to account for the latest advances in nuclear physics. 
Several statistical models have been
developed accounting for the entire distribution of nuclei (Hempel \& Schaffner-Bielich 2010, 
Raduta \& Gulminelli 2010, Buyukcizmeci \etal 2013) 
in contrast to the approximation of a single representative
heavy nucleus and $\alpha$-particles within the
classical EoS (Lattimer \& Swesty 1991, Shen \etal 1998). 
Except in some small density and
temperature regions, the effects on global thermodynamics remains
small and the effect on core collapse simulations is modest (Steiner \etal 2013). 
The effect on the neutrino interactions and spectra could be more important (Arcones \etal 2008).\\
For progenitors with masses above $\sim25 M_{\rm sol}$, the temperatures and densities reached during core collapse can become so high that a traditional
description in terms of electrons, nuclei, and nucleons is no longer
adequate. Several EoS have been developed incorporating additional
particles, such as hyperons and pions (Oertel \etal 2012, Gulminelli \etal 2013, Banik
\etal 2014). Among the consequences, the time necessary to collapse to a black
hole is reduced (Peres \etal 2013).

\subsection{Confirmed axisymmetric explosions from first principles for a wider set of progenitors}

A series of paper by Suwa \etal (2010), M\"uller \etal (2012a, 2012b, 2013), and Bruenn \etal (2013) confirmed the viability of the neutrino-driven explosion mechanism in the axisymmetric hypothesis. Successful explosions have been obtained from first principles for progenitor masses ranging from 8.1 to $27 M_{\rm sol}$, both by the MPA group and the Oak Ridge group. One may regret that the explosion times and energies measured by these two groups are not yet fully compatible with each other, and neutrino transport  still relies on some numerical approximations, but these recent studies are still very encouraging steps towards a consensus. 
Both groups use a ray-by-ray method which solves the Boltzmann equation in each radial direction assuming an axisymmetric distribution of neutrinos along each radial ray. It is interesting to note that the Princeton group did not obtain an explosion using a different code where neutrino transport is approximated with a multi-group, flux-limited diffusion method (Dolence \etal 2014). \\
Although these explosions are an immense success in view of the decades of failed numerical attempts, the mechanism cannot be qualified as robust enough yet, since the final explosion energy seems significantly too low compared to observations. It is not clear whether the solution to this puzzle will come from improving the modeling of progenitor models, or including 3D effects from rotation and magnetic fields, or something else.

\subsection{The diversity of explosion paths}

Despite the universality of the Chandrasekhar mass defining the initial conditions of core collapse, the past five years have taught us that the explosion process is sensitive to the radial structure of the envelope surrounding the central core.

\subsubsection{The parameter $\chi$ and the relative roles of neutrino-driven convection vs SASI}

The numerical simulations performed at MPA and ORNL confirmed the diversity of hydrodynamical processes ruling the multidimensional evolution during the last hundreds of milliseconds before the explosion. 
Neutrino-driven buoyancy seems to govern the post-shock dynamics of the progenitor of 11.2 $M_{\rm sol}$. The global oscillations of SASI are clearly dominant for the 27 $M_{\rm sol}$ progenitor (M\"uller \etal 2012b), while both instabilities seem to be entangled in the evolution the 15$M_{\rm sol}$ progenitor. Such differences can be understood from the comparison of the advection time and the buoyancy times in these systems. A short advection time is both favorable to SASI (Foglizzo \etal 2007, Scheck \etal 2008), and stabilizing for neutrino-driven convection (Foglizzo \etal 2006). By adjusting the neutrino luminosity and the rate of dissociation at the shock, Fernandez \etal (2014) were able to characterize each instability separately and compare the properties of the buoyant bubbles and turbulence induced before the explosion.  \\
Besides a few examples in the parameter space, we do not have a clear picture of the dependence of the $\chi$ parameter on the main sequence mass yet. It should be noted that the development of SASI is not a sufficient criterion for a successful explosion. The axisymmetric simulation of a $25 M_{\rm sol}$ progenitor is a clear  example of strong SASI activity without an explosion (Hanke \etal 2013).

\subsubsection{The compactness parameter $\xi_{2.5}$ and black hole vs neutron star formation}

A systematic study performed by Ugliano \etal(2012) used some educated prescription to account for multidimensional evolution using 1D simulations over a wide range of progenitor masses from 10 to 40 solar masses. The outcome of the core collapse can either be a supernova explosion with the formation of a neutron star, a failed supernova with the formation of a black hole, or a supernova with the delayed formation of a black hole through the fall back of enough ejecta. The diversity of mass loss processes during the life of a star results in a non-monotonous relation between its mass at the moment of core-collapse and its mass on the main sequence. In addition to stressing this labeling confusion, the study of Ugliano \etal (2012) showed that the production of a neutron star or a black is also very variable for progenitors more massive than 15 $M_{\rm sol}$. An interesting indicator is based on the compactness parameter $\xi_{2.5}$ defined by O'Connor \& Ott (2011), measuring the radius of the innermost 2.5$M_{\rm sol}$: 
\begin{eqnarray}
\xi_{2.5}\equiv {M/M_{\rm sol}\over R(M)/1000{\rm km}}.
\end{eqnarray}
Although this indicator is not strictly deterministic, a threshold $\xi_{2.5}>0.3$ seems very favorable to black hole formation in the calculations of Ugliano \etal (2012). From a series of 101 axisymmetric simulations from 10.8 to 75$M_{\rm sol}$ with IDSA neutrino transport, Nakamura \etal (2014a) found that a high compactness parameter corresponds to a later explosion time, a stronger neutrino luminosity and explosion energy, and a higher Nickel yield.

\subsubsection{The possible influence of pre-collapse convective asymmetries}

The structure of the progenitor may also influence the asymmetry of the explosion mechanism through the convective inhomogeneities associated with thermonuclear burning in the Silicon and Oxygen shells (Arnett \& Meakin 2011). 
The quantitative impact of this process has been estimated by Couch \& Ott (2013), who showed an example where these 
asymmetries had an effect comparable to a $2\%$ increase of the neutrino luminosity and could be enough to turn a failed explosion into a successful one. Elaborating in this direction, the recent study of M\"uller \& Janka (2014b) pointed out that the most efficient effect of pre-collapse asymmetries comes from inhomogeneities with the largest angular scale $l=1,2$.

\subsection{The unexpected difficulties of 3D explosions}

\subsubsection{3D explosions are not easier than 2D}

Using a simplified model of neutrino interactions with cooling and heating functions, Nordhaus \etal (2010b) had claimed that the neutrino luminosity needed to obtain an explosion was lower in 3D than in 2D, in a comparable proportion as measured by Murphy \& Burrows (2008) between 2D and 1D. This raised the hope that the weak explosions obtained in 2D axisymmetric simulations could become more robust once computer ressources would allow first principle calculations including neutrino transport in 3D. A careful check by Hanke \etal (2012) and Couch (2013) showed that the neutrino luminosity threshold in the idealized setup studied by Nordhaus \etal (2010b) is actually very similar in 3D and 2D. The misleading results of Nordhaus \etal (2010b) happened to be due to some numerical errors in the treatment of gravity in their CASTRO code, later corrected and acknowledged by Burrows \etal (2012). Despite these corrections, the convergence between the different codes is not yet fully satisfactory (Dolence \etal 2013).  \\
Using the IDSA approximation of neutrino transport, Takiwaki \etal (2014) studied the evolution of a 11.2 $M_{\rm sol}$ progenitor and observed that the explosion time is shorter and the explosion is more vigorous in 2D than in 3D. Similar conclusions were reached by Couch \& O'Connor (2014) with a multi species leakage scheme for progenitors of 15 and 27 $M_{\rm sol}$. Let us note that the stochastic nature of the dynamical evolution and the prohibitive cost of 3D simulations make it difficult to achieve numerical convergence. Beside the possible influence of numerical artifacts (\eg Ott \etal 2013 compared to Abdikamalov \etal 2014), part of the difference between 2D and 3D simulations could be related to a more efficient energy cascade from large scales to small scales in 3D than in 2D (Hanke \etal 2012, Couch 2013, Couch \& O'Connor 2014).

\subsubsection{SASI does exist in 3D models of core collapse}

According to the mechanism proposed by Guilet \etal (2010) and checked on the axisymmetric simulations of Fernandez \& Thompson (2009), the saturation amplitude of SASI oscillations is not expected to be very different in 2D and 3D. The waves of entropy and vorticity created by the oscillations of the shock are expected to lose their large scale $l=1$ coherence when disrupted by the parasitic growth of Rayleigh-Taylor and Kelvin-Helmholtz instabilities. This description, however, ignores the interaction of SASI with small scale turbulence, whose properties may differ in 2D and 3D.
Even though the first 3D simulations performed by Blondin \& Mezzacappa (2007) in an adiabatic approximation showed well defined spiral shock patterns on a large scale, subsequent 3D simulations with less idealized setups suggested weaker amplitudes than in 2D (Iwakami \etal 2008, Burrows \etal 2012). The clear dominance of non-oscillatory convective motions in some simulations even led Burrows \etal (2012) to draw general conclusions such that SASI would be an artefact of simplified physics which does not exist in realistic 2D simulations, and even less in 3D. This reasoning missed the diversity of explosion paths later explored in a parametric manner by Fernandez \etal (2014) in 2D or Iwakami \etal (2014a) in 3D, and exemplified by the axisymmetric simulation by M\"uller \etal (2012) of the 27 $M_{\rm sol}$ progenitor. The evolution of this progenitor was computed in 3D from first principles by Hanke \etal (2013) who confirmed the presence of the SASI mode with an amplitude which can exceed even the amplitude observed in 2D. Nevertheless, the explosion did not take place during the first 400ms of evolution, whereas it did explode in 2D on this timescale. It is interesting to note the sensitivity of the dynamics to small differences in the equation of state and possibly GR effects between the axisymmetric results of Hanke \etal (2013) and those of M\"uller \etal (2012), where less than 200ms were sufficient for an explosion.

\subsection{Rotation and magnetic fields}

The angular momentum of the stellar core and its magnetic field are two physical ingredients which definitely affect the birth properties of the neutron star and may also influence the explosion mechanism.
 
 \subsubsection{SASI effect on the pulsar spin}

The spectacular conclusions regarding the pulsar spin obtained by Blondin \& Mezzacappa (2007) with a 3D adiabatic setup were naturally followed by more realistic 3D simulations. Fernandez (2010) was able to confirm the angular momentum budget when the progenitor does not rotate, with a setup similar to Blondin \& Shaw (2007) but in 3D. By contrast, the spiral mode did not appear easily in the non-rotating simulation of Iwakami \etal (2009) who considered a stationary accretion flow incorporating the buoyancy effects due to neutrino heating and a more realistic equation of state. Significantly slower pulsar spin periods of the order of 500-1000ms were estimated by Wongwathanarat \etal (2010) who considered the collapse of a 15$M_{\rm sol}$ progenitor with neutrino cooling and heating, self gravity and general relativistic corrections of the monopole, and a Yin-Yang grid. Similar skeptical conclusions were reached by Rantsiou \etal (2011), who pointed out that the mass cut during the explosion process might not coincide with the separation of positive and negative angular momentum produced by SASI.
A larger sample of non-rotating progenitors was considered by Wongwathanarat \etal (2013), including a 20$M_{\rm sol}$ progenitor, and measured final spin periods of 100ms to 8000ms. They noted that the kick and spin do not show any obvious correlation regarding their magnitude or direction. Their analysis clarified an important difference between the kick and spin mechanisms: the spin up takes place during the phase of accretion onto the neutron star surface because it is dominated by the direct accretion of angular momentum. By contrast, the kick can grow on a longer time scale, even when accretion has stopped, through the action of the gravitational force. Using analytic arguments, Guilet \& Fernandez (2014) estimated the maximum spin of a neutron star born from a non-rotating progenitor by measuring the amount of angular momentum which can be stored in the saturated spiral mode of SASI. They obtained spin periods compatible to those measured in the simulations of Wongwathanarat \etal (2013).

 \subsubsection{Rotation effects on the shock dynamics}

The effect of a moderate rotation on the dynamics of the collapse has received little attention so far, despite the fact that interesting correlations between the kick and spin directions would be expected in this regime. Indeed, the growth of a prograde spiral SASI mode is expected from the early numerical simulations of Blondin \& Mezzacappa (2007), and the linear stability analysis of Yamasaki \& Foglizzo (2008) indicated that this effect of differential rotation can be significant even if the centrifugal force is weak. The 3D simulations of stationary accretion by Iwakami \etal (2009) confirmed that the amplitude of the prograde spiral mode of SASI is enhanced by rotation. Iwakami \etal (2014b) gradually increased the angular momentum of the infalling gas to measure the influence of rotation on the explosion threshold. According to the few points in their Fig.~1, the change in the threshold of neutrino luminosity required to obtain an explosion seems to be a linear function of the angular momentum with a luminosity variation of $\sim 10\%$ for an angular momentum of the order of $\sim5\times10^{15}$cm$^2$s$^{-1}$. More simulations would be necessary to check whether this effect really scales linearly with the rotation rate as expected. According to a tentative linear extrapolation, rotation effects with an angular momentum of $\sim 10^{15}$cm$^2$s$^{-1}$ could induce a $2\%$ shift of the neutrino luminosity, \ie a similar order-of-magnitude effects as the pre-collapse asymmetries described by Couch \& Ott (2013). For reference, $10^{15}$cm$^2$s$^{-1}$ is the equatorial angular momentum at the surface of a pulsar with a radius of 10km and a spin period of 6ms. The uncertain effect of mixing instabilities inside the neutron star remains to be evaluated. Compared to the models of stellar evolution by Heger \etal (2005), a value of  $10^{15}$cm$^2$s$^{-1}$ is twice as large as the estimated angular momentum at 2000km with magnetic fields ($\sim 5\times10^{14}$ cm$^2$s$^{-1}$), and 20 times smaller than without magnetic fields ($\sim2\times10^{16}$ cm$^2$s$^{-1}$).\\
Nakamura \etal (2014b) considered the collapse of the stellar core of a $15 M_{\rm sol}$ progenitor with a shell type rotation profile: $\Omega(r)=\Omega_0 /(1+r^2/R_0^2)$ with $R_0=2\times10^{8}$cm and $\Omega_0=0$, $0.1\pi$, and $0.5\pi$ rad s$^{-1}$ corresponding to an initial angular momentum at 2000km of $0$, $6\times 10^{15}$cm$^2$s$^{-1}$, and $3\times 10^{16}$cm$^2$s$^{-1}$. They measured a reduction of the threshold of neutrino luminosity of the order of $10\%$ for $\Omega_0=0.1\pi$, which is consistent with the results of Iwakami \etal (2014b). They emphasized that the direction of the explosion is preferentially perpendicular to the spin axis but did not take into account the asymmetry of neutrino emission induced by rotation, which may favour the axial direction.

\subsubsection{The growth of magnetic energy without differential rotation}

Endeve \etal (2010, 2012) used an adiabatic setup to show that even when the initial rotation of the core is neglected, the magnetic energy can grow from the turbulent motions induced by SASI to reach $10^{14}$G at the surface of the neutron star. The growth of this magnetic field, however, did not significantly affect the dynamics of the shock. Obergaulinger \etal (2014) addressed the same question in a more realistic setup including neutrino-driven convection and a M1 approximation of neutrino transport. They confirmed the lack of significant effects on the shock evolution except if the field is initially very strong. They found evidence for an accumulation of Alfven waves at the Alfven point as described by Guilet \etal (2011), but did not find a significant contribution to the field amplification or heating.

\subsubsection{Magnetic field amplification with fast rotation}

Fast differential rotation provides a large energy reservoir for magnetic
field amplification, such that magnetic effects cannot be neglected beyond
a certain angular momentum. The shearing of a poloidal magnetic field into
a toroidal one provides a linear amplification, which would be relevant
only if the initial poloidal magnetic field were large enough. If the
initial poloidal magnetic field is weak, the exponential growth of the
magnetorotational instability (MRI) is a more promising mechanism of
magnetic field amplification (Akiyama \etal 2003). A meaningful
description of this process requires 3D simulations with a very high
resolution, due to the very short wavelength of the MRI growing on a weak
initial field. The prohibitive computing time required has led researchers
to use axisymmetric simulations with artificial assumptions such as a
strong initial poloidal field mimicking the outcome of some amplification
processes (\eg Moiseenko \etal 2006, Burrows \etal 2007, Takiwaki \etal
2009, Takiwaki \& Kotake 2011). In this framework, they obtained powerful
magnetorotational explosions, with jets launched along the polar axis due
to the winding of the initial poloidal magnetic into a very strong
($\gtrsim 10^{15}\,{\rm G}$) toroidal field. The progress of computational
power has opened new perspectives to address this problem in 3D. Winteler
\etal (2012) suggested that such magnetorotational explosions could be an
important source of $r$-process elements. The 3D simulations of M\"osta
\etal (2014) highlighted the role of a non-axisymmetric instability of the
toroidal magnetic field which can break the jet structure and prevent such
magnetorotational explosions, a non-runaway expansion of the shock being
observed instead.\\
In parallel progress is being made in understanding the magnetic field
amplification due to the MRI in the protoneutron star. 
Buoyancy driven by radial gradients of entropy
and lepton fraction has a strong impact on the dynamics, though the
thermal and lepton number diffusion due to neutrinos allows for fast MRI
growth, as shown by the linear analysis of Masada \etal (2006, 2007). 
The study of the non-linear phase of the MRI requires 3D
simulations, which have been undertaken in local and semi-global models
that describe a small fraction of the protoneutron star. Taking into
account global gradients responsible for buoyancy has, however, proven
difficult in this framework because of boundary effects (Obergaulinger
\etal 2009) or coarse radial resolution (Masada \etal 2014).
A local model in the Boussinesq
approximation allowed Guilet \& M\"uller (2015)
to estimate that a stable stratification decreases the efficiency of magnetic field amplification only slightly, 
but significantly impacts the structure of the magnetic field. \\
Neutrino radiation, neglected in most numerical simulations, can slow
down the MRI growth through an effective neutrino viscosity deep inside the
proto-neutron star or through a neutrino drag near its surface (Guilet \etal 2014). 
Further local simulations will be useful to study the MRI in this new growth regime, and
ultimately global simulations (only achieved in 2D so far (Sawai \etal
2013, Sawai \& Yamada 2014)) will be needed to assess the magnetic field
geometry and its impact on the explosion.

\section{AN EXPERIMENTAL APPROACH TO SUPERNOVA DYNAMICS\label{sect_fountain}}

\begin{figure}
\begin{center}
\includegraphics[width=\columnwidth]{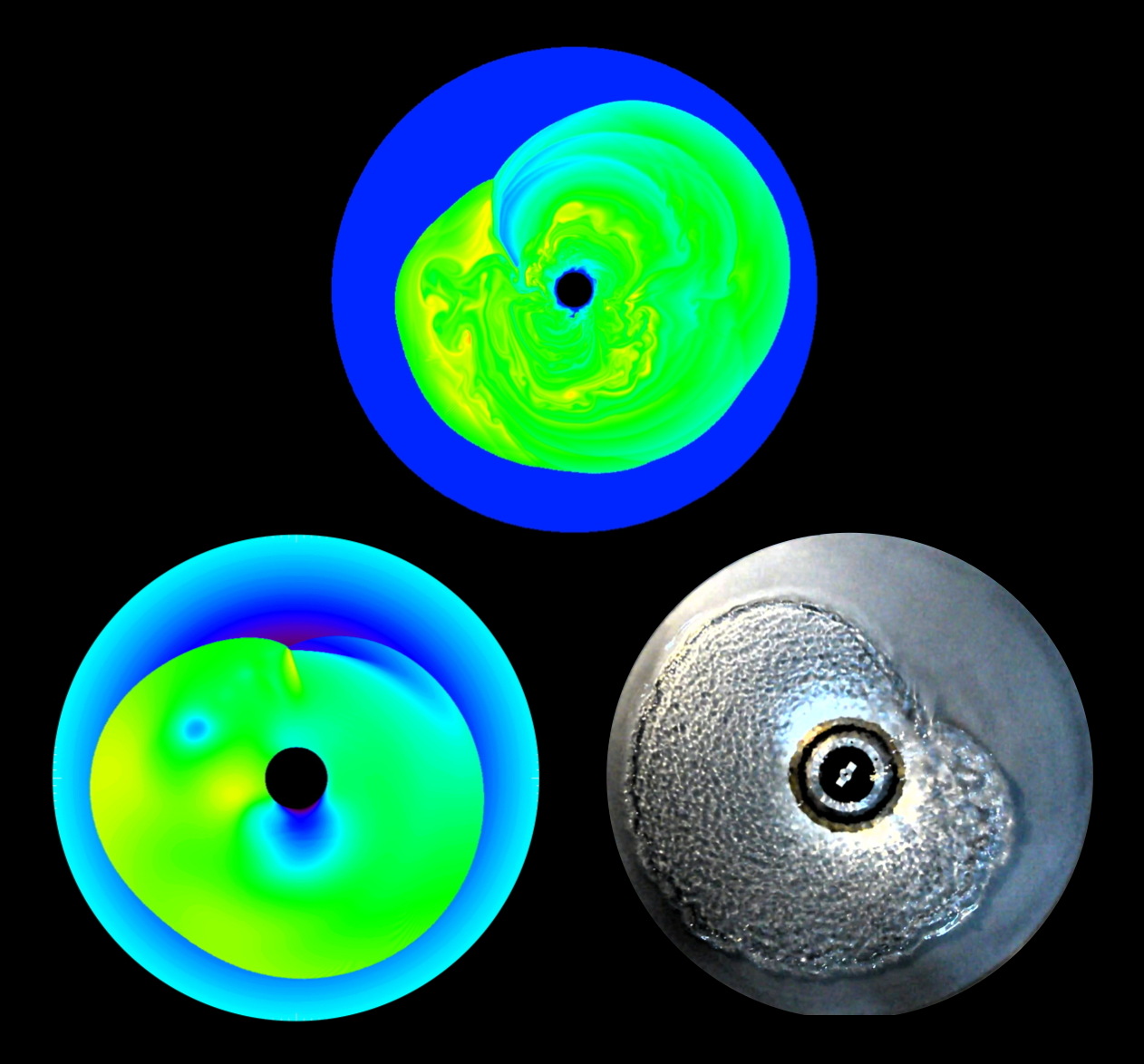}
\caption{The shape of the rotating hydraulic jump observed in the non-linear regime in the SWASI experiment (right) is similar to the shape observed in the shallow water approximation (left) and the shape of the SASI in the numerical simulations of cylindrical gas accretion with local neutrino cooling (top).}\label{Fig_tryptique}
\end{center}
\end{figure}

In order to make the SASI phenomenon more intuitive, Foglizzo \etal (2012) designed a shallow water analogue of the gas motion in the equatorial plane of the stellar core. Surface gravity waves in water play the role of
acoustic waves in the stellar gas and participate in an unstable cycle with vorticity perturbations. Solving numerically the 2D system of shallow water equations revealed the similarity between SASI and its shallow water analogue. The absence of buoyancy effects in this shallow water formulation is theoretically instructive, as will be shown further. The first laboratory analogue of SASI dynamics was obtained using a water fountain with a diameter of 60cm. This new research tool can help us become more familiar with SASI, in the same manner as we are familiar with the flapping of a flag or the convection of a fluid heated from below.

\subsection{From astrophysical complexity to the simplicity of a water fountain}
 
\subsubsection{From gas to water dynamics: the shallow water analogy}

In the simple fountain built by Foglizzo \etal (2012), the dynamics of water looks very similar to the gas dynamics induced by SASI in the equatorial plane of the star, on a scale one million times smaller and an oscillation period one hundred times slower than in the astrophysical flow. On the one hand, the fury of extreme gravity accelerating the stellar gas to one tenth of the speed of light in less than half a second. Iron nuclei broken apart upon compression by a shock wave which moves back and forth with a 30ms period. On the other hand, water flowing at room pressure and temperature in a meter size fountain, giving rise to the back and forth motions of a surface wave with a 3 second period. The striking resemblance illustrated in Fig.~\ref{Fig_tryptique} is rooted in the similarity between the Euler equations describing an inviscid adiabatic gas and the Saint Venant equations describing shallow water.
The conservation of mass, momentum and energy for a gas with an adiabatic index $\gamma$, density $\rho$, pressure $P$, entropy $S$, velocity $v$, and sound speed $c_{\rm s}^2=\gamma P/\rho$ is written as follows:
\begin{eqnarray}
{\p \rho\over\p t}+\nabla\cdot (\rho v)&=&0,\\
{\p v\over\p t}+(\nabla\times v)\times v+\nabla \left\lbrack{v^2\over 2}+{c_{\rm s}^2\over\gamma-1}+\Phi\right\rbrack&=&c_{\rm s}^2\nabla S,\label{Euler}\\
{\p S\over\p t}+v\cdot\nabla S&=&0.
\end{eqnarray}
The pressure gradient $\nabla P$ in the Euler equation (\ref{Euler}) has been decomposed into a enthalpy gradient $\rho\nabla (c_{\rm s}^2/(\gamma-1))$ and an entropy gradient $\rho c_{\rm s}^2\nabla S$. \\
The Saint Venant set of equations describing the shallow layer of water with depth $H$ flowing with a velocity $v$ over a surface $z=H_\Phi(r)$ is the following:
\begin{eqnarray}
{\p H\over\p t}+\nabla\cdot (H v)&=&0,\label{mass}\\
{\p v\over\p t}+(\nabla \times v)\times v+\nabla \left\lbrack{v^2\over 2}+c_{\rm w}^2+\Phi\right\rbrack&=&0.\label{stvenant}
\end{eqnarray}
The local gravity $g=981{\rm cm\;s}^{-2}$ in the laboratory infuences the speed $c_{\rm w}$ of surface waves in the experiment and the effective potential $\Phi\equiv gH_\Phi$, which mimicks the Newtonian gravity of the central object with a hyperbolic shape:
\begin{eqnarray}
c_{\rm w}^2&\equiv&gH,\\
H_\Phi(r)&\equiv& -{(5.6{\rm cm})^2\over r}.
\end{eqnarray}
The Froude number defined by ${\rm Fr}=|v|/c_{\rm w}$ plays the same role as the Mach number ${\cal M}=|v|/c_{\rm s}$ in a gas. The pressure variation $\Delta P$ induced by the weight of water $\Delta P=\rho_{\rm w}g (z-H_\Phi(r))$ has a mean value noted $P=\rho_{\rm w}g H/2$ related to the local speed of surface waves $c_{\rm w}$:
\begin{eqnarray}
P=\rho_{\rm w} {c_{\rm w}^2\over 2}.
\end{eqnarray}
This relation is analogous to the relation $P=\rho c_{\rm s}^2/\gamma$ for a gas with an adiabatic index $\gamma=2$.

\subsubsection{The absence of buoyancy effects in shallow water}

Besides the particular value of the adiabatic index, the main difference between the two sets of equations comes from the absence of buoyancy effects in shallow water. This limitation precludes the study of buoyancy driven instabilities such as neutrino driven convection and their interesting interaction with SASI. This limitation also 
ignores the formation of buoyant entropy bubbles from shock oscillations driven by SASI (\eg Fernandez \etal 2014), and the dominant role of buoyancy driven parasitic instabilities to saturate the amplitude of SASI (Guilet \etal 2010). Even in the inviscid limit the dynamical system described by the shallow water equations can only describe the interaction of surface waves and vorticity perturbations, whereas the Euler equations describe the interaction of acoustic waves, entropy and vorticity perturbations. In a positive sense, the shallow water dynamical system offers an opportunity to assess the importance of buoyancy effects on the dynamics of the shock. 

\subsubsection{Inner boundary condition}

\begin{figure}
\begin{center}
\includegraphics[width=\columnwidth]{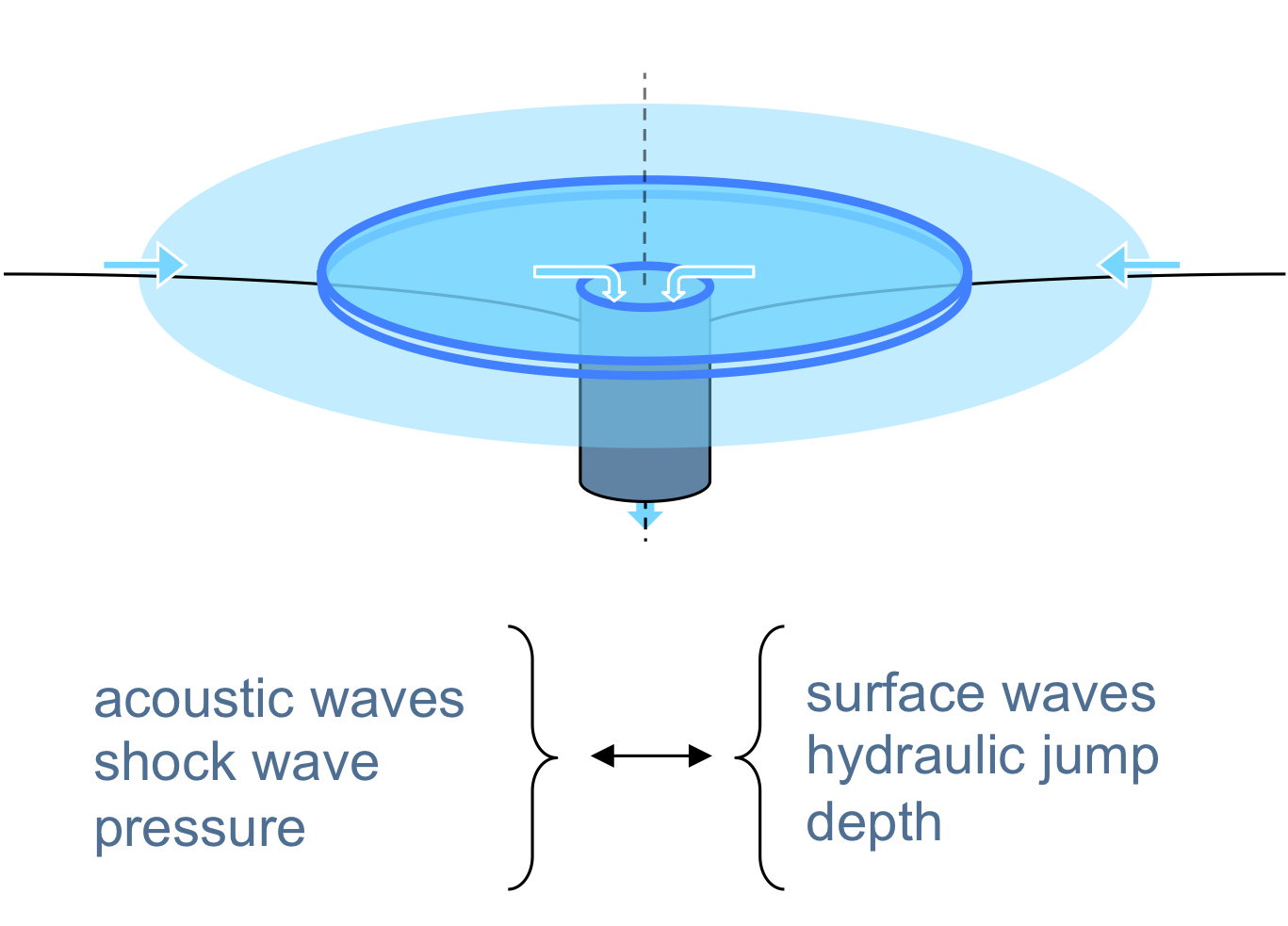}
\caption{Water is injected inward from a circular slit, in a uniform and stationary manner. A circular hydraulic jump is produced by the vertical surface of the cylinder at the center. Water is evacuated by spilling over the upper edge of this cylinder. }\label{Fig_experiment}
\end{center}
\end{figure}

The inner boundary condition in the experiment faces the same difficulty as the adiabatic simulation of Blondin \& Mezzacappa (2007): without any non-adiabatic process to shrink the volume of the accreted fluid on the surface of the neutron star, the fluid has to be continuously extracted through the inner boundary. This is done in the experiment by allowing water to spill over the upper edge of the inner cylinder (Fig.~\ref{Fig_experiment}). As long as the radius $R_{\rm ns}$ of the cylinder is large enough to allow the free fall of water inside it, this inner boundary condition can be translated into the condition that the radial component of the flow velocity $|v_{\rm ns}|$ reaches the local wave velocity $c_{\rm w}(R_{\rm ns})$ associated to the depth $H_{\rm edge}$ of water above the upper edge of the inner cylinder. Denoting by $Q$ the flow rate,
\begin{eqnarray}
2\pi R_{\rm ns} H_{\rm edge}|v_{\rm ns}| &=& Q,\\
v_{\rm ns}^2&=&gH_{\rm edge}.
\end{eqnarray}

\subsubsection{SWASI, a shallow water analogue of SASI}

In view of the similarity of the equations, the formalism of advected-acoustic cycles studied by Foglizzo (2001, 2002, 2009) can be translated into a linear coupling process between surface gravity waves and vorticity perturbations, to produce a cycle between the hydraulic jump and the inner boundary. How would the absence of feedback from the advection of entropy perturbation affect the stability of this cycle in the shallow water analogue? The perturbative study and numerical simulations of the shallow water equations performed by Foglizzo \etal (2012) confirmed the existence of an instability very similar to SASI, named SWASI as a shallow water analogue of a shock instability. Despite the simplicity of the shallow water approximation, an excellent agreement was found between the oscillation period measured in the experiment and the one predicted by the shallow water equations.

\subsection{Experimental limitations}

\subsubsection{Input parameters}

The variable parameters of the experiment are the flow rate $0<Q<4{\rm L}/{\rm s}$, the thickness of the injection slit $0.5<H_{\rm inn}\le2$mm, the height $-5.6{\rm cm}<z_{\rm ns}<0$ and diameter $8\le D_{\rm ns}\le16$cm of the inner cylinder. The radius of injection is set to $33$cm. The minimum size of the injection slit is limited by its building accuracy.
The diameter of the inner cylinder sets the maximum flow rate above which the inner boundary should be modeled as a drowned boundary rather than a critical point. A central spinner, made of a light object in free rotation along a central vertical axis, is in passive contact with the surface of the water in the vicinity of the inner cylinder. Its rotating motion helps visualize the angular momentum in the vicinity of the neutron star.
In the new version described in Sect.~\ref{sect_SWASI_rot}, the rotation of the full experiment is motorized and can be adjusted electronically. 

\subsubsection{Hydraulic jumps vs shock waves}

The hydraulic jump in the experiment is a reverse version of the familiar circular hydraulic jump observed in kitchen sinks. The water flowing inward faster than the speed of waves (${\rm Fr}>1$) is abruptly decelerated to slower velocities (${\rm Fr}<1$) into a thicker layer of water. Mass flux is exactly conserved across this transition, and momentum flux is approximately conserved due to the minor drag on the fountain surface. Jump conditions similar to the Rankine-Hugoniot conditions are deduced from these two conservation laws. The conservation of the energy flux is not expected since the radial flow of water is described by only two variables, $v$ and $H$, without any equivalent of the entropy variable in a gas. The same is true for isothermal shocks in astrophysics, where the excess energy is usually assumed to be radiated away. In the hydraulic jump with a moderate Froude number, excess energy is viscously dissipated within the thickness of the jump by one or several horizontal stationary rollers. Other processes such as wave emission, the formation of bubbles or splashes can participate in the evacuation of the excess energy at a higher Froude number (\eg Chanson 2009).
The 2D formulation of the shallow water equations does not resolve the internal structure of the hydraulic jump and simply assumes that energy is instantly dissipated across an idealized discontinuity. \\
The description of hydraulic jump as a discontinuity is also unable to describe the limit of Froude numbers approaching unity $1<{\rm Fr}<2$, where the radial extent of the hydraulic jump increases with a succession of oscillations referred to as an undular hydraulic jump. According to the classification of Chow (1973), the range of Froude numbers most favorable to the simplest shallow water description of the experiment is $4<{\rm Fr}<9$.

\subsubsection{Viscosity effects}

The strongest effect of water viscosity takes place ahead of the hydraulic jump, in the shallowest region where the vertical shear is responsible for a significant viscous drag. In the experiment presented in Foglizzo \etal (2012), this viscous drag has been measured and modeled as a laminar drag in Eq.~(\ref{stvenant}):
\begin{eqnarray}
{\p v\over\p t}+(\nabla \times v)\times v+\nabla \left\lbrack{v^2\over 2}+c^2+\Phi\right\rbrack&=&
\alpha\nu_{\rm w} {v\over H^2},\label{stvenant2}
\end{eqnarray}
where $\nu_{\rm w}\sim 0.01{\rm cm}^2{\rm s}^{-1}$ is the kinematic viscosity of water and $\alpha\sim3$ is a dimensionless number measured in the experiment.\\
Fortunately this viscous drag is very small in the deeper region after the hydraulic jump and does not significantly affect the dynamics of the interaction between surface waves and vorticity perturbations. In order to compensate for the viscous drag ahead of the hydraulic jump, the water can be injected from the outer boundary with a higher velocity than the local free fall velocity, so that the velocity immediately before the jump $R_{\rm jp}\sim 20$cm coincides with the local inviscid free fall velocity $v_{\rm ff}(R_{\rm jp})\equiv (2gH_\Phi(R_{\rm jp}))^{1/2}$.\\
Increasing the injection flow rate increases the development of turbulence in the fountain, with an expected transition for a Reynolds number of the order of ${\rm Re}\sim2000$. 
The viscosity of water seems too weak to affect the dynamics of SASI in the laminar regime, but a turbulent viscosity could significantly affect the efficiency of the cycle between surface waves and vorticity perturbations.

\subsubsection{Uncertain vertical profile of velocity}

The inviscid formulation of shallow water equations corresponds to an idealized situation where the velocity $v$ is uniform in the vertical direction. The non-uniform vertical velocity profile viscously induced by the no-slip condition introduces a modification of the quadratic terms such as $(\nabla\times v)\times v$ and $\nabla v^2/2$ in the shallow water equations, since the vertically averaged value of $v^2$ is no longer equal to the square of the vertical average of $v$. Correction coefficients known as the Coriolis and Boussinesq coefficients could account for this effect if the vertical profile of velocity were known, for example in the idealized cases of a laminar flow or in fully developed turbulence. 

\subsubsection{A 2D view of the equatorial plane of the stellar core}

The shallow water experiment is intrinsically limited to 2D since the radial attraction of the central neutron star is mimicked from a projection of the vertical gravity of the laboratory onto the inclined surface of the fountain. The shallow water equations could, however, be solved in 3D if necessary, for example to assess the importance of buoyancy by comparison with 3D simulations of SASI. 
 
\subsubsection{From shallow water equations to the experiment and back}

From the point of view of understanding the essence of the physics of SASI using analogies, the idealized set of inviscid shallow water equations is the simplest formulation with the most direct astrophysical connections. These equations can be studied analytically and solved numerically without using the experimental fountain. The water experiment is more complex because of viscosity effects which are not trivially captured in models and have no direct astrophysical interpretation such as the development of a boundary layer, the viscous drag, Coriolis and Boussinesq coefficients, the radial extension of the hydraulic jump and its connection to bubble entrainment, splashes or unsteady behavior unrelated to SASI. A careful interpretation of experimental results is needed to identify the results which are governed by the simplest shallow water equations. The experiment is complementary to simulations because its limitations are different from the numerical ones, such as the difficulty of first order numerical convergence due to the presence of a shock (Sato \etal 2009). This difficulty is inherent to any shock capturing method. The simplicity of use of the experiment makes it very convenient to rapidly explore a large  parameter space and identify new phenomena. From an astrophysical perspective, the immediate next step consists in plugging the experimental parameters of this new phenomenon in a numerical simulation of the simplest shallow water equations with a viscous drag. If the phenomenon is confirmed numerically, its sensitivity to the viscosity is tested by performing new shallow water simulations in the inviscid limit. The next step could be a 2D equatorial simulation of adiabatic gas dynamics to incorporate buoyancy effects, and a series of simulations to successively incorporate a hard inner boundary with a cooling function, neutrino driven convection with a heating function, a third dimension, an equation of state for nuclear matter, neutrino transport and interactions, relativistic corrections, a progenitor profile for the initial conditions and an inner boundary condition which allows the contraction of the proto-neutron star. It is important to bear in mind the extreme simplicity of the shallow water setup compared to the most advanced simulations such as Hanke \etal (2013). It is equally important to bear in mind that even the simplest setup like the shallow water formulation contains interesting hydrodynamical puzzles and surprises.

\subsection{Building up our physical intuition about SASI}

\subsubsection{SASI is not familiar yet}

Part of the apparent complexity in the physical problem of core collapse comes from hydrodynamical instabilities such as neutrino driven convection and SASI, which break the spherical symmetry. Most physicists are familiar with the convective instability, even though its interaction with a shock wave and an advection flow is far from trivial. By contrast, SASI is much less familiar to astrophysicists. Its advective-acoustic mechanism relates to aeroacoustic instabilities in ramjets (Abouseif \etal 1984), rocket motors (Mettenleiter \etal 2000), and can be traced back to the whistling of a kettle (Chanaud \& Powel 1965). In astrophysics the instability of Bondi-Hoyle-Lyttleton accretion also relates to this category of unstable advective-acoustic cycles (Foglizzo \etal 2005), but the conical geometry of the shock make it less intuitive than SASI and this problem has been much less studied than stellar core collapse. Characterizing the fundamental properties of SASI in their simplest formulation makes it possible to enlighten more complex simulations (\eg Blondin \& Mezzacappa 2007).

\subsubsection{The linear coupling between vorticity perturbations and waves}

\begin{figure}
\begin{center}
\includegraphics[width=\columnwidth]{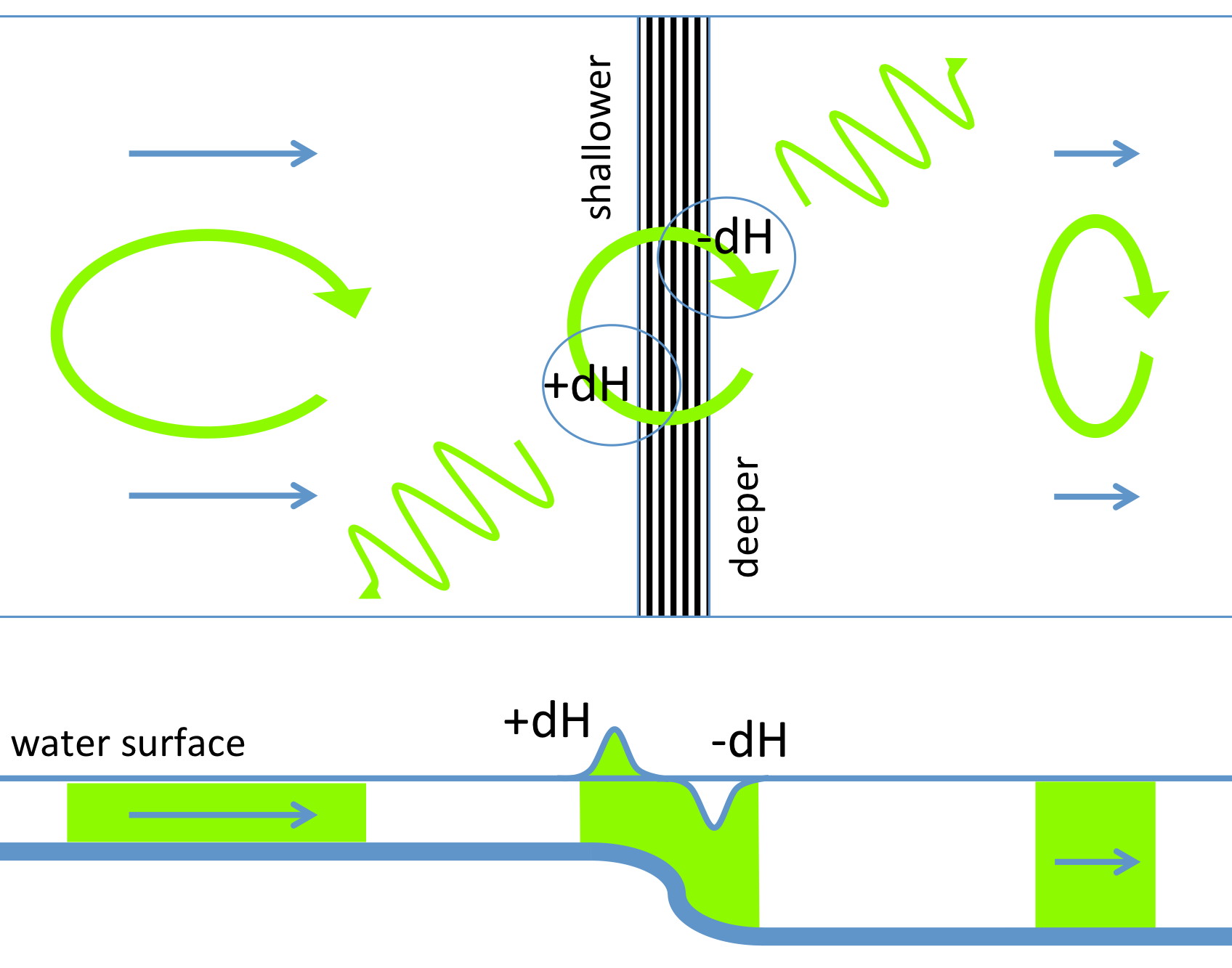}
\caption{The coupling between vorticity perturbations and surface gravity waves in shallow water can help understand the coupling between vorticity perturbations and acoustic waves in a gas. The vorticity perturbation is shown in green at three successive positions in the upper illustration, viewed from above. The change in water elevation $\pm\delta H$ produced by the vortical motion over the gradient of depth is viewed horizontally in the lower illustration. It is a source of surface gravity waves.}\label{Fig_coupling}
\end{center}
\end{figure}

The translation of gas dynamics into shallow water dynamics is in many respects a fruitful exercise, which can produce physical pictures sometimes more intuitive with water than with a gas although they are absolutely equivalent in the end. 
Let us describe the coupling between vorticity waves and surface gravity waves in an inhomogeneous flow. A vortical perturbation can be advected with the flow at uniform velocity without ever affecting the pressure equilibrium. Let us imagine that this perturbation is slowly advected in a direction where the depth of the shallow layer of fluid increases, as illustrated in Fig.~\ref{Fig_coupling}. Because we have an intuitive representation that the surface of slow water (\ie ${\rm Fr}\ll 1$) remains horizontal even when the depth increases abruptly (\eg $\p H_\Phi/\p r \gg1$), we are able to guess that the vortical motion must perturb this flat surface by transporting some water from the deep regions to shallower regions and conversely. 
We also have an intuitive understanding that the perturbed water surface will return to horizontal through the propagation of surface waves. The same intuitive reasoning can be translated into an isentropic or isothermal gas decelerated by some external potential. 
A similar approach was used by Foglizzo \& Tagger (2000) to explain the feedback from the advection of an entropy perturbation across the adiabatic compression produced by an external potential. It should be noted that these two feedback processes are fundamentally different but since entropy and vorticity perturbations are advected with the same velocity, the feedback they produce adds up in a coherent advective-acoustic cycle studied analytically by Foglizzo (2009) and numerically by Sato \etal (2009). If a magnetic field were present, vorticity and entropy perturbations would no longer propagate at the same velocity and the advective-acoustic cycle would be transformed into five distinct MHD cycles (Guilet \& Foglizzo 2010).

\subsubsection{Open questions accessible to the SWASI fountain}

The SWASI fountain is currently being used as a complementary tool to numerical simulations to address the following questions:
\par (i) Comparing the growth rate of SWASI in the experiment to the growth rate deduced from the shallow water equations with various prescriptions for the viscous drag and the turbulent viscosity can teach us about the sensitivity of the SWASI mechanism to experimental limitations.

\par (ii) Understanding the interaction of SWASI with the turbulence produced by the vertical shear may shed light on the physics of SASI and its interaction with the turbulence induced by parasitic instabilities or by neutrino driven convection.

\par (iii) Measuring the experimental conditions for the symmetry breaking between the clockwise and anti-clockwise modes of SWASI can help us understand the nature of this non-linear process.

\par (iv) Understanding the saturation amplitude of SASI is essential to predict which type of progenitors could lead to the strongest SASI oscillations. Is the growth of parasitic instabilities proposed by Guilet \etal (2010) always the dominant saturation mechanism? Testing the saturation mechanism of SWASI in the shallow water setup may be simpler than SASI in a gas because of the action of a single parasitic instability (Kelvin-Helmholtz) associated with the fragmentation of vorticity waves.

\subsubsection{A new experiment including the angular momentum of the stellar core\label{sect_SWASI_rot}}

A new experiment has been built at CEA Saclay with similar dimensions as the one published by Foglizzo \etal (2012). The main difference is the capability to motorize the rotation of the full experiment so that water can be injected with a non-zero angular momentum. This device should be able to experimentally address the effect of rotation on the growth of the prograde spiral mode (Yamasaki \& Foglizzo 2008) and its saturation amplitude. Based on the adverse rotation of the central spinner observed by Foglizzo \etal (2012) in a flow without angular momentum, one can expect to experimentally demonstrate the competition between the angular momentum of the background flow and the adverse angular momentum in the region close to the neutron star (Blondin \& Mezzacappa 2007).

\subsection{A tool for public outreach}

\begin{figure}
\begin{center}
\includegraphics[width=\columnwidth]{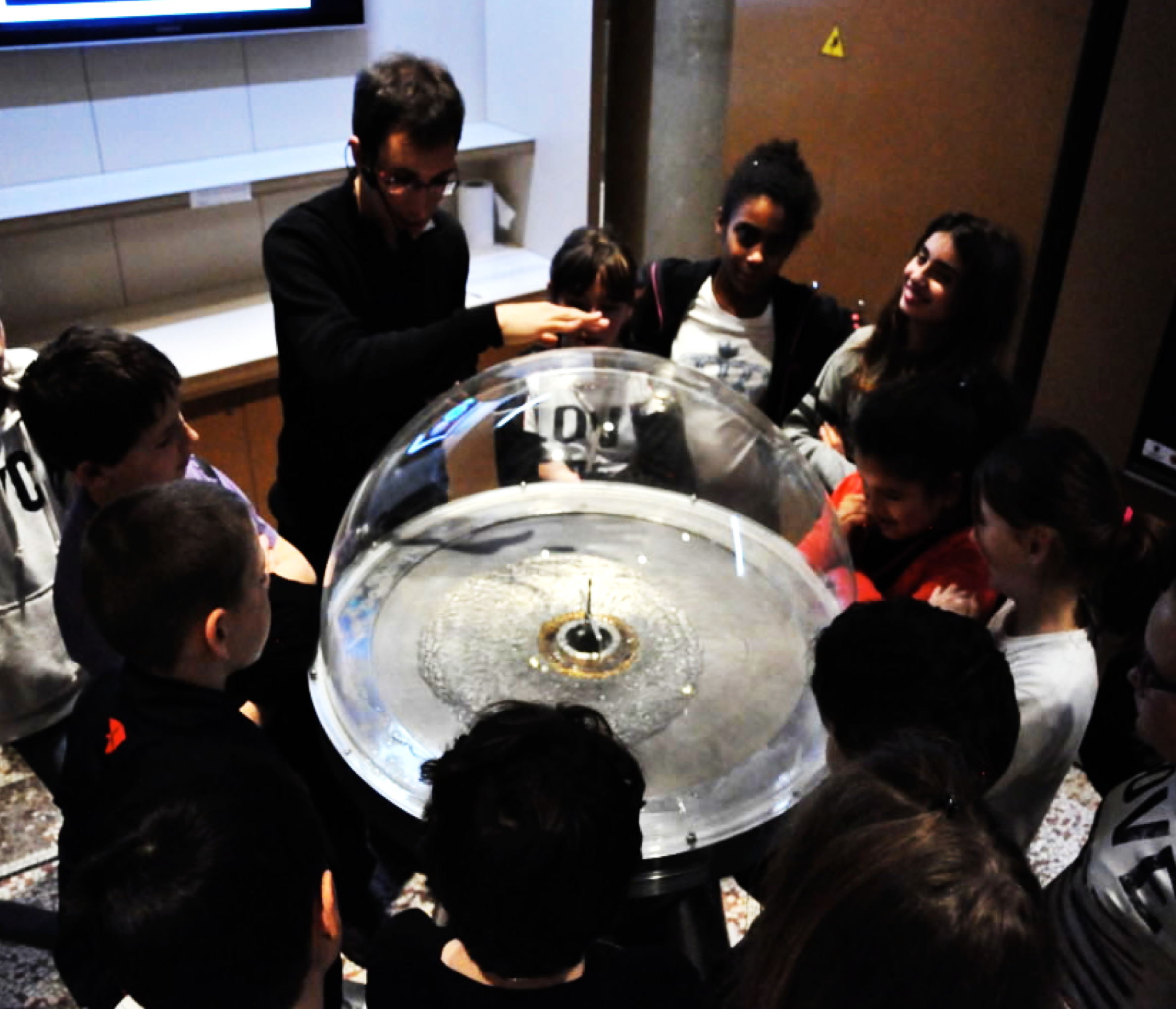}
\caption{In 2014 the SWASI experiment was presented by the SN2NS collaboration (here R. Kazeroni) in the Paris Science Museum to explain supernova theory to the public.}\label{Fig_museum}
\end{center}
\end{figure}

A simplified version of the SWASI experiment was exposed to the public of the Science Museum in Paris, from December 2013 to February 2014 (Fig.~\ref{Fig_museum}). The fountain was accompanied by a series of explanatory posters, a video about cosmic nucleosynthesis and a bouncing ball experiment.
Two to four presentations of 40mn were proposed every day by the 12 researchers of the SN2NS collaboration for an average audience of 15 people aged 7 and above. It is remarkable that some of the most recent discoveries in supernova physics could be transmitted to more than 2,000 visitors during this period. The interest of the public was clear enough for the museum to decide to incorporate this supernova fountain into their permanent collection as from 2015. This supernova fountain was awarded the 2014 prize for scientific communication from the French Ministry of Research and Higher Education.

\subsubsection{From bouncing balls to fluid mechanics}

As an introduction to the fluid experiment, the public was invited to comment on a side experiment were 
two bouncing balls fall vertically, both guided by a wire attached between the ceiling and the ground. 
The diameter of the lower bouncing ball is bigger ($\sim  8$cm) than the upper one ($\sim5$cm). Used with a single ball, this experiment first illustrates the question of elastic vs inelastic bounce. When the two balls are dropped simultaneously the spectacular transmission of energy and momentum from the lower ball to the upper one is always a surprise to the public witnessing the ejection of the upper ball to the ceiling. This experiment is a good introduction to the simplest  and naive solution to the problem of core-collapse. The SWASI experiment takes the public two steps further, both in the direction of fluid mechanics and with the additional degree of freedom allowed by transverse motions. The radial flow of water in the experiment is presented as a view of the dynamics in the equatorial plane of the stellar core.

\subsubsection{Familiarity with water and kitchen sink hydraulic jumps}

One key aspect of the public success of the SWASI fountain is the simplicity of the setup, and the familiarity of the public with water. The main action on the fountain was a change in flow rate from 0 to 1L/s which showed the radial injection of water, and the formation of a circular hydraulic jump at a radius of $\sim20$cm. The comparison with the circular hydraulic jump visible in every kitchen sink helps illustrate the analogy with shocks, and identify the subsonic and supersonic parts of the flow. The changes in composition are also introduced during this phase of stationary accretion: while the supersonic gas is made of iron nuclei, the post-shock gas is dissociated into protons, neutrons and electrons, and the neutron star is essentially made of neutrons with neutrinos diffusing out. It is important to introduce the neutrinos as a key actor of the supernova mechanism early enough. This is one of the difficulties of the presentations since this particle is generally unknown to the public, and absent from the analogous experiment. The post-shock gas is described as dense enough to capture a fraction of the outgoing neutrino flux, but insufficiently to revive the shock when distributed spherically.
Unlike the stable jump in a kitchen sink, the hydraulic jump in the supernova fountain moves in an irresistible manner. Its motion is as inevitable as the fall of a pen in vertical equilibrium on a table. Like the falling pen the initial symmetry of the equilibrium is broken in a random direction. The breaking of the circular symmetry appears with the development of $m=1$ sloshing oscillations along a random direction with a 3s period. As the oscillation of the hydraulic jump reaches larger amplitude, a second symmetry breaking leads to the rotation of the spiral wave in a random direction, clockwise or anti-clockwise. The central spinner raises the curiosity of the public by rotating erratically in the direction opposite to the hydraulic jump.

\subsubsection{Capturing astrophysical processes accessible on a human scale}

The notions of time-scale and length-scale are introduced with the fountain, applying the factor one million to scale down the neutron star radius, the shock radius, the size of the iron core and the size of a supergiant. The factor $\sim$hundred in timescale is applied to speed up the observed dynamics of the hydraulic jump. As the neutron star is identified with the surface of the inner cylinder, the extreme properties of these compact stars are stressed: nuclear densities, extreme spin rates, mysterious velocities up to 1000km/s.
The scaling factors enable the public to capture representations of extreme astrophysical processes on a human scale. 

\subsubsection{Public access to the physics of explosion threshold, kick and spin}

The connection between the dynamics of the fountain and stellar explosions is explained along three observations:
\par(i) the explosion threshold is understood by observing the radial extent of the post-shock material which can intercept neutrinos in the direction of largest shock displacement. Neutrino absorption will be enhanced in this direction compared to the spherically symmetric case.
\par(ii) the spectacular pulsar kicks up to 1000km/s can be reached  if the explosion takes place in an asymmetric manner: this is a consequence of the conservation of linear momentum, which can easily be experienced by feeling the recoil when throwing a heavy object 
\par(iii) the significant  spin up of the neutron star is illustrated by the motion of the central spinner. The conservation of angular momentum is invoked to accept that the spinner and the hydraulic jump rotate in opposite directions, noting that the injected flow is purely radial.

\section{CONCLUSION\label{sect_conclusion}}

The development of 3D numerical simulations over the last two years has allowed a more detailed comparison between theoretical models and observations. The most recent results suggest that the general framework of neutrino driven asymmetric explosions is consistent with several observations when the neutrino luminosity is adjusted to trigger the explosion. Among them, the kick of neutron stars (from Scheck \etal 2004 to Wongwathanarat \etal 2013) and the chemical mixing in the stellar envelope (from Kifonidis \etal 2006 to Wongwathanarat \etal 2014). The distribution of $^{44}$Ti clumps in the supernova remnant Cassiopea A (Grefenstette \etal 2014) seems to confirm the theoretical prediction by Wongwathanarat \etal (2013). On the downside, we should remember that we do not yet understand how to produce a robust supernova explosion from first principles. The last five years have shown several signs of convergence between a growing number of research groups. Part of the difficulties may come from
the cost of checking the numerical convergence of 3D simulations in a dynamical system involving stochasticity and shocks in addition to the sensitivity to different numerical approximations. \\
The dependence of the explosion mechanism on the pre-collapse structure of the stellar core has opened up the parameter space of initial conditions and stressed the reliance on the implementations of non-spherical effects in stellar evolution codes such as rotation, magnetic fields and convective asymmetries. The additional effect of binary interaction during the lifetime of a massive star may open the parameter space even more.
The choice of initial conditions for 3D numerical simulations must be guided by a better understanding of the domain of dominance of the different hydrodynamical instabilities at work. Similarly the only way to assess if the results from a simulation are general or coincidental is the detailed understanding of the mechanisms ruling their evolution. 
The shallow water analogy was developed as a complementary tool to improve our intuition about the dynamics of SASI. Its potential for public outreach has already been tested on the public with encouraging results.\\
In the near future, supernova experts are impatient to discover possible clues through the properties of the compact object hiding behind the asymmetric ejecta of SN1987A. Even more direct constraints could come from the detection of neutrinos and gravitational waves from the next galactic supernova. There is ample work for theoreticians before these much awaited events occur.

\begin{acknowledgements}

This work is part of the ANR funded project SN2NS ANR-10-BLAN-0503. TF acknowledges insightful discussions with Thomas Janka, Bernhard M\"uller, Rodrigo Fernandez, S\'ebastien Fromang, Elias Khan, Christian Ott, Kei Kotake, Shoichi Yamada, Chris Fryer, John Blondin, Luc Dessart, Jeremiah Murphy, Steve Bruenn, Adam Burrows, Wakana Iwakami, Selma de Mink, Yudai Suwa and Sean Couch. The Science Museum ``Palais de la D\'ecouverte" in Paris is thanked for the help from Sylvain Lefavrais, Kamil Fadel, Guillaume Trap and Denis Savoie. Elias Khan and Michael Urban are thanked for their help with the presentations at the museum. TF is grateful to the technical team who helped to design and build the new versions of the supernova fountain: Patrice Charon and Michael Massinger at CEA/IRFU/SIS, Emmanuel Gr\'egoire at CEA/IRFU/SAp, Patrick Oriol and Pascal Schummer at SDMS. TF also thanks the dedicated students who helped to explore the dynamics of SWASI, Adrien Kuntz, David Martin, Audrey Ch\^atain, Clement Royer, Alexandra Bouvot.

\end{acknowledgements}
%

\end{document}